Higher Order Squeezing and Higher Order Subpoissonian Photon Statistics in Intermediate States

Amit Verma[1] and Anirban Pathak[2]

Department of Physics and Materials Science and Engineering,

Jaypee Institute of Information Technology,

A-10, Sector-62, Noida, UP-201307,

India

**Abstract**

Recently simpler criteria for the Hong-Mandel higher order squeezing (HOS) and higher order subpoissonian photon statistics (HOSPS) are provided by us [Phys. Lett. A **374** (2010) 1009]. Here we have used these simplified criteria to study the possibilities of observing HOSPS and HOS in different intermediate states, such as generalized binomial state, hypergeometric state, negative binomial state and photon added coherent state. It is shown that these states may satisfy the condition of HOS and HOSPS. It is also shown that the depth and region of nonclassicality can be controlled by controlling various parameters related to intermediate states. Further, we have analyzed the mutual relationship between different signatures of higher order nonclassicality with reference to these intermediate states. We have observed that the generalized binomial state may show signature of HOSPS in absence of HOS. Earlier we have shown that NLVSS shows HOS in absence of HOSPS. Consequently it is established that the HOSPS and HOS of same order are independent phenomenon.

# 1 Introduction

A state which does not have any classical analogue is known as nonclassical state. To be precise, when the Glauber Sudarshan P-function of a radiation field become negative or more singular than a delta function then the radiation field is said to be nonclassical. In these situations quasi probability distribution P is not accepted as classical probability and thus we can not obtain an analogous classical state. For example, squeezed state and antibunched state are well known nonclassical states. These two lowest order nonclassical states have been studied since long but the interest in higher order nonclassical states is relatively new. Possibilities of observing higher order nonclassicalities in different physical systems have been investigated in recent past [1-16]. For example, i) higher order squeezed state of Hong Mandel type [1-4] (HOS), ii) higher order squeezed state of Hillery type [5,6], iii) higher order subpoissonian photon statistics (HOSPS) [7-9] and iv) higher order antibunched state (HOA) [10-16] are recently studied in different physical systems. But the general nature of higher order nonclassicality and the mutual relation between these higher order nonclassical states required more attention [1]. Further, recently we have shown that the higher order antibunching can be observed in most of the intermediate states [16]. We have also observed HOS and HOSPS in binomial state, HOS in non-linear vacuum squeezed state (NLVSS) and HOSPS in non-linear excited squeezed state (NLESS) [1]. These observations lead to an immediate question: Do the other intermediate states satisfy the criteria of HOS and HOSPS? The present work aims to answer these question and study the mutual relationship between the criteria of higher order nonclassicalities. Another reason behind the study of higher order nonclassical properties of intermediate states lies in the fact that the most of the interesting recent developments in quantum optics have arisen through the nonclassical properties of the radiation field only. But the majority of these studies are focused on lowest order nonclassical effects.

Commonly, second order moment (standard deviation) of an observable is considered to be the most natural measure of quantum fluctuation [17] associated with that observable and the reduction of quantum fluctuation below the coherent state (poissonian state) level corresponds to lowest order nonclassical state. For example, an electromagnetic field is said to be electrically squeezed field if uncertainties in the quadrature phase observable $X$ reduces below the coherent state level (i.e. $(\Delta X)^2 < \frac{1}{2}$) and antibunching is defined as a phenomenon in which the fluctuations in photon number reduces below the Poisson level (i.e. $(\Delta N)^2 < \langle N\rangle$) [18,19]. This condition can be generalized and we can say that a quantum mechanical state $|\psi\rangle$ has $nth$ order nonclassicality with respect to an arbitrary quantum mechanical operator $A$ if the $n$th order moment of $A$ in that state reduces below the value of the $nth$ order moment of $A$ in a poissonian state, i.e. the condition of $nth$ order nonclassicality with respect to the operator $A$ is

$$(\Delta A)^n_{|\psi\rangle} < (\Delta A)^n_{|poissonian\rangle}, \tag{1}$$

[1]amit.verma@jiit.ac.in

[2]anirbanpathak@yahoo.co.in

where $(\Delta A)^n$ is the $n$th order moment defined as

$$\langle(\Delta A)^n\rangle = \sum_{r=0}^{n} {}^nC_r(-1)^r\overline{A^r}\ \bar{A}^{n-r}. \tag{2}$$

If $A$ is a field operator then it can be expressed as a function of creation and annihilation operators $a$ and $a^\dagger$ and consequently further simplification of (1) is possible by using the identity

$$\langle:\left(A(a,a^\dagger)\right)^k:\rangle_{|poisonian\rangle} = \langle\left(A(a,a^\dagger)\right)\rangle^k_{|poisonian\rangle} \tag{3}$$

where, the notation $:(A(a,a^\dagger)^k:$ is simply a binomial expansion in which powers of the $a^\dagger$ are always kept to the left of the powers of the $a$. It is clear from (2) that the problem of finding out the $nth$ order moment of the operator $A$ essentially reduces to a problem of operator ordering (normal ordering) of $A^r$. Here, we would like to note that we observe the lowest order nonclassicality for $n=2$. And in this particular case ($n=2$) we obtain the condition of squeezing of electric field, if $A = X = \frac{1}{\sqrt{2}}(a+a^\dagger)$ and obtain the condition of antibunching if $A = N = a^\dagger a$. In our recent work [1], we have generalized the idea of these well known lower order nonclassical effects using this particular notion[3] of higher order nonclassicality and the normal ordered form of $X^r$ and $N^r$.

In section 2 and 3 we have briefly presented the simplified criteria [1]of HOS and HOSPS respectively. In section 4, it is shown that the HOS and HOSPS may be observed in generalized binomial state, hypergeometric state, negative binomial state and photon added coherent state. Role of various parameters in controlling the depth and region of nonclassicality is also discussed. Finally in section 5 we conclude.

# 2 Simplified criterion of higher order squeezing

In our recent work [1] we have used an operator ordering theorem introduced by us in [20] to obtain the simplified condition for higher order squeezing (HOS) of Hong Mandel type [2, 4]. In that work we had claimed that our formalism considerably simplifies the calculation of HOS. To establish that we will investigate the possibilities of observing HOS in intermediate states which are known to be higher order nonclassical as it has been already shown that these states show higher order antibunching [16]. Now the condition for HOS for the usual quadrature variable $X$ is obtained as

$$\langle(\Delta X)^n\rangle < \frac{1}{2^{\frac{n}{2}}}t_n = \frac{1}{2^{\frac{n}{2}}}(n-1)!! = \left(\frac{1}{2}\right)_{\frac{n}{2}} \tag{4}$$

or,

$$\sum_{r=0}^{n}\sum_{i=0}^{\frac{r}{2}}\sum_{k=0}^{r-2i}(-1)^r\frac{1}{2^{\frac{n}{2}}}t_{2i}\,{}^{r-2i}C_k\,{}^nC_r\,{}^rC_{2i}\langle a^\dagger+a\rangle^{n-r}\langle a^{\dagger k}a^{r-2i-k}\rangle < \left(\frac{1}{2}\right)_{\frac{n}{2}}. \tag{5}$$

Starting from the generalized notion of higher order nonclassicality (1) this closed form expression of Hong-Mandel squeezing is obtained in [1]. This condition of HOS significantly reduces the calculational difficulties. To be precise, to study the possibility of HOS for an arbitrary quantum state $|\psi\rangle$ we just need to calculate $\langle a^\dagger+a\rangle$ and $\langle a^{\dagger k}a^{r-2i-k}\rangle$. Calculation of this expectation values are simple. For example, if we can expand the arbitrary state $|\psi\rangle$ in the number state basis as

$$|\psi\rangle = \sum_{j=0}^{N}C_j|j\rangle, \tag{6}$$

then we can easily obtain,

$$\langle\psi|a^{\dagger k}a^{r-2i-k}|\psi\rangle = \sum_{j=0}^{N-Max[k,r-2i-k]}C^*_{j+k}C_{j+r-2i-k}\frac{1}{j!}\left((j+k+r-2i)!(j+k)!\right)^{\frac{1}{2}} \tag{7}$$

where Max yields the largest element from the list in its argument and

$$\langle a^\dagger+a\rangle = \sum_{m=0}^{N-1}\sqrt{(m+1}\left(C_mC^*_{m+1}+C^*_mC_{m+1}\right). \tag{8}$$

[3]According to this notion of higher order squeezing Hillery type amplitude powered squeezing is lower order squeezing of nonlinear bosonic operators ($Y_1$and $Y_2$). This is so because the amplitude powered squeezing is described by the reduction (with respect to the poissonian state) of second order moment of the corresponding quadrature variable. One can easily extend the existing notion of Hillery type squeezing and obtain a new kind of higher order nonclassicality, namely, Hong Mandel type squeezing of Hillery type operator.

Therefore,

$$
\begin{array}{rcl}
\langle(\Delta X)^n\rangle & = & \sum_{r=0}^{n}\sum_{i=0}^{\frac{r}{2}}\sum_{k=0}^{r-2i}(-1)^r\frac{1}{2^{\frac{n}{2}}}t_{2i}\,{}^{r-2i}C_k\,{}^nC_r\,{}^rC_{2i} \\
 & \times & \left(\sum_{m=0}^{N-1}\sqrt{(m+1}\left(C_mC_{m+1}^*+C_m^*C_{m+1}\right)\right)^{n-r} \\
 & \times & \sum_{j=0}^{N-Max[k,\,r-2i-k]}C_{j+k}^*C_{j+r-2i-k}\frac{1}{j!}\left((j+k+r-2i)!(j+k)!\right)^{\frac{1}{2}}.
\end{array} \tag{9}
$$

In general, if we know the effect of $a^s$ on the state $|\Psi\rangle$ and the orthogonality conditions $\langle\Psi'|\Psi\rangle$ then we can easily find out $\langle(\Delta X)^n\rangle$. Further, since (9) is a $C$-number equation, analytical tools like MAPPLE and MATHEMATICA can also be used to study the possibilities of observing higher order squeezing (or higher order nonclassicality in general). This point will be more clear in section 4, where we will provide specific examples. Here we would like to note that we can normalize (4) and rewrite the condition of HOS as

$$
S_{HM}(n)=\frac{\langle(\Delta X)^n-\left(\frac{1}{2}\right)_{\frac{n}{2}}}{\left(\frac{1}{2}\right)_{\frac{n}{2}}}<0 \tag{10}
$$

where the subscript $HM$ stands for Hong Mandel.

# 3 Simplified criterion of higher order subpoissonian photon statistics

In analogy to the procedure followed to derive the simplified criterion of Hong-Mandel type of higher order squeezing we can also derive a criterion for HOSPS from the generalized expression (1) of higher order nonclassicality. The criterion is presented in detail in [1]. The criterion is

$$
\langle(\Delta N)^n\rangle=\sum_{r=0}^{n}{}^nC_r(-1)^r\bar{N}^r\bar{N}^{n-r}=\sum_{r=0}^{n}\sum_{k=1}^{r}S_2(r,k)\,{}^nC_r(-1)^r\langle N^{(k)}\rangle\langle N\rangle^{n-r}<\langle(\Delta N)^n\rangle_{|poissonain\rangle}
$$

or,

$$
d_h(n-1)=\sum_{r=0}^{n}\sum_{k=1}^{r}S_2(r,k)\,{}^nC_r(-1)^r\langle N^{(k)}\rangle\langle N\rangle^{n-r}-\sum_{r=0}^{n}\sum_{k=1}^{r}S_2(r,k)\,{}^nC_r(-1)^r\langle N\rangle^{k+n-r}<0. \tag{11}
$$

where $S_2(r,k)$ is the Stirling number of second kind $N^{(k)}=a^{\dagger k}a^k$ is the $kth$ factorial moment of the number operator $N$. The negativity of $d_h(n-1)$ will mean $(n-1)th$ order subpoissonian photon statistics. This condition of HOSPS is equivalent to the condition of HOSPS obtained by Mishra-Prakash [9].

The main aim of the present paper is to study the possibilities of observing HOS and HOSPS in intermediate states. But we will require the information about HOA for the study of mutual relationship between different kind of higher order nonclassical effects. The signature of $l$th order antibunching is reflected through the negativity of

$$
d(l)=\langle N^{(l)}\rangle-\langle N\rangle^l. \tag{12}
$$

The closed form analytic expressions of $d(l)$ for the intermediate states of interest may be found in our earlier publication [16]. Here we have graphically presented $d(l)$ to understand the relationship among HOA, HOS and HOSPS.

# 4 HOS and HOSPS in intermediate states

An intermediate state is a quantum state which reduces to two or more distinguishably different states in different limits. In 1985, such a state was first time introduced by Sloter *et al.* [21]. In brief, they introduced Binomial state (BS) as a state which is intermediate between the most nonclassical number state $|n\rangle$ and the most classical coherent state $|\alpha\rangle$. Since the introduction of BS as an intermediate state, different properties (such as, antibunching, squeezing and higher order squeezing) of binomial states have been studied [22-26]. This trend of search for nonclassicality in Binomial state, continued in ninetees and in one hand, several versions of generalized BS has been proposed [22-24] and in the other hand people went beyond binomial states and proposed several other form of intermediate states (such as, odd excited binomial state [26], hypergeometric state [27], negative hypergeometric state [28], reciprocal binomial state [29], and photon added coherent state [30] etc.). The studies in the 90s were mainly limited to theoretical predictions but the recent developments in the experimental techniques made it possible to verify some of those theoretical predictions. For example, we can note that, as early as in 1991 Agarwal and Tara [30] introduced photon added coherent state as

$$
|\alpha,m\rangle=\frac{a^{\dagger m}|\alpha\rangle}{\langle\alpha|a^m a^{\dagger m}|\alpha\rangle^{1/2}}, \tag{13}
$$

(where $m$ is an integer and $|\alpha\rangle$ is coherent state) but the experimental generation of the state has happened only in recent past when Zavatta, Viciani and Bellini [31] succeeded to produce it in 2004. It is easy to observe that this is an intermediate state, since it reduces to coherent state in the limit $m \to 0$ and to number state in the limit $\alpha \to 0$. This state can be viewed as a coherent state in which additional $m$ photon are added. The photon number distribution of all the above mentioned states are different but all these states belong to a common family of states called intermediate state. It is also been found that most of these intermediate states show antibunching, squeezing, higher order squeezing, Subpoissonian photon statistics etc.

In the present work, we have shown that the intermediate states (e.g. GBS, NBS, HS and PACS) satisfy the simplified criteria (10 and 11) derived in our recent work [1]. This paper provides observation of higher order nonclassical characteristics HOS and HOSPS in intermediate states (GBS, NBS, PACS and HS). In the following subsections we will study possibilities of observing HOS and HOSPS in GBS, NBS, PACS and HS. Let us start with the GBS.

## 4.1 Generalized Binomial State

As we have mentioned earlier there are different form of generalized binomial states [22-24], in the present section we have chosen generalized binomial state introduced by Roy and Roy [23] for our study. Roy and Roy have introduced the generalized binomial state as

$$|N,\alpha,\beta\rangle = \sum_{n=0}^{N} \sqrt{\omega(n,N,\alpha,\beta)}|n\rangle \tag{14}$$

where,

$$\omega(n,N,\alpha,\beta) = \frac{N!}{(\alpha+\beta+2)_N}\frac{(\alpha+1)_n(\beta+1)_{N-n}}{n!(N-n)!} \tag{15}$$

where $(x)_r$ is conventional Pochhammer symbol and $\alpha,\beta > -1$, $n = 0,1,....,N$. Now with the help of properties of Pochhammer symbol and operator algebra we can obtain following relations:

$$\begin{array}{lcl} a\,|N,\alpha,\beta\rangle & = & \left[\frac{N\,(\alpha+1)}{(\alpha+\beta+2)}\right]^{1/2}\sum_{n'=0}^{N-1}\sqrt{\omega(n',N-1,\alpha+1,\beta)}\,|n'\rangle \\ a^2\,|N,\alpha,\beta\rangle & = & \left[\frac{N(N-1)(\alpha+1)(\alpha+2)}{(\alpha+\beta+2)(\alpha+\beta+3)}\right]^{1/2}\sum_{n'=0}^{N-2}\sqrt{\omega(n',N-2,\alpha+2,\beta)}\,|n'\rangle \\ \vdots & \vdots & \vdots \\ a^l\,|N,\alpha,\beta\rangle & = & \left[\frac{N!(\alpha+l)!(\alpha+\beta+1)!}{(N-l)!\alpha!(\alpha+\beta+l+1)!}\right]^{1/2}\sum_{n'=0}^{N-l}\sqrt{\omega(n',N-l,\alpha+l,\beta)}\,|n'\rangle\,, \end{array} \tag{16}$$

$$\langle\beta,\alpha,N|\,a^{\dagger k} = \langle n'|\left[\frac{N!(\alpha+k)!(\alpha+\beta+1)!}{(N-k)!\alpha!(\alpha+\beta+k+1)!}\right]^{1/2}\sum_{n'=0}^{N-k}\sqrt{\omega(n',N-k,\alpha+k,\beta)}, \tag{17}$$

$$\langle a\rangle = \langle a^\dagger\rangle \;=\; \left[\frac{N\,(\alpha+1)}{(\alpha+\beta+2)}\right]^{1/2}\sum_{n'=0}^{N-1}\sqrt{\omega(n',N-1,\alpha+1,\beta)\omega(n',N,\alpha,\beta)}\;, \tag{18}$$

$$\left\langle a^{\dagger k}a^k\right\rangle = \left[\frac{N!(\alpha+k)!(\alpha+\beta+1)!}{(N-k)!\alpha!(\alpha+\beta+k+1)!}\right] \tag{19}$$

and

$$\begin{array}{lcl}\left\langle a^{\dagger k}a^l\right\rangle & = & \left[\frac{N!^2(\alpha+k)!(\alpha+l)!(\alpha+\beta+1)!^2}{(N-k)!(N-l)!\alpha!^2(\alpha+\beta+k+1)!(\alpha+\beta+l+1)!}\right]^{1/2} \\ & & \sum_{n'=0}^{N-Max[k,l]}\sqrt{\omega(n',N-k,\alpha+k,\beta)}\sqrt{\omega(n',N-l,\alpha+l,\beta)}\end{array}. \tag{20}$$

Therefore,

$$\begin{array}{lcl} d_h(n-1)_{GBS} & = & \sum_{r=0}^{n}\sum_{k=1}^{r}\left[S_2(r,k)\,{}^nC_r(-1)^r\left\{\left\langle a^\dagger a\right\rangle\right\}^{n-r}\left\{\left\langle a^{\dagger k}a^k\right\rangle-\left\langle a^\dagger a\right\rangle^k\right\}\right] \\ & = & \sum_{r=0}^{n}\sum_{k=1}^{r}\left[S_2(r,k)\,{}^nC_r(-1)^r\left\{Mp\right\}^{n-r}\left\{\frac{N!(\alpha+k)!(\alpha+\beta+1)!}{(N-k)!\alpha!(\alpha+\beta+k+1)!}-\left(\frac{N\,(\alpha+1)}{(\alpha+\beta+2)}\right)^k\right\}\right], \end{array} \tag{21}$$

and

$$\begin{array}{lcl} S_{HM}(n)_{GBS} & = & \frac{1}{\left(\frac{1}{2}\right)_{\frac{n}{2}}}\sum_{r=0}^{n}\sum_{i=0}^{\frac{r}{2}}\sum_{k=0}^{r-2i}(-1)^r\frac{1}{2^{\frac{n}{2}}}t_{2i}\,{}^{r-2i}C_k\,{}^nC_r\,{}^rC_{2i}\langle a^\dagger+a\rangle^{n-r}\langle a^{\dagger k}a^{r-2i-k}\rangle-1 \\ & = & \frac{1}{\left(\frac{1}{2}\right)_{\frac{n}{2}}}\left[\sum_{r=0}^{n}\sum_{i=0}^{\frac{r}{2}}\sum_{k=0}^{r-2i}(-1)^r\frac{1}{2^{\frac{n}{2}}}t_{2i}\,{}^{r-2i}C_k\,{}^nC_r\,{}^rC_{2i}\left[2\left[\frac{N\,(\alpha+1)}{(\alpha+\beta+2)}\right]^{1/2}\right.\right. \\ & & \left.\sum_{n'=0}^{N-1}\sqrt{\omega(n',N-1,\alpha+1,\beta)\,\omega(n',N,\alpha,\beta)}\right]\left[\frac{N!^2(\alpha+k)!(\alpha+r-2i-k)!(\alpha+\beta+1)!^2}{(N-k)!(N-r+2i+k)!\alpha!^2(\alpha+\beta+k+1)!(\alpha+\beta+r-2i-k+1)!}\right] \\ & & \left.\sum_{n'=0}^{N-Max[k,r-2i-k]}\sqrt{\omega(n',N-k,\alpha+k,\beta)}\sqrt{\omega(n',N-r+2i+k,\alpha+r-2i-k,\beta)}\right]-1 \end{array}. \tag{22}$$

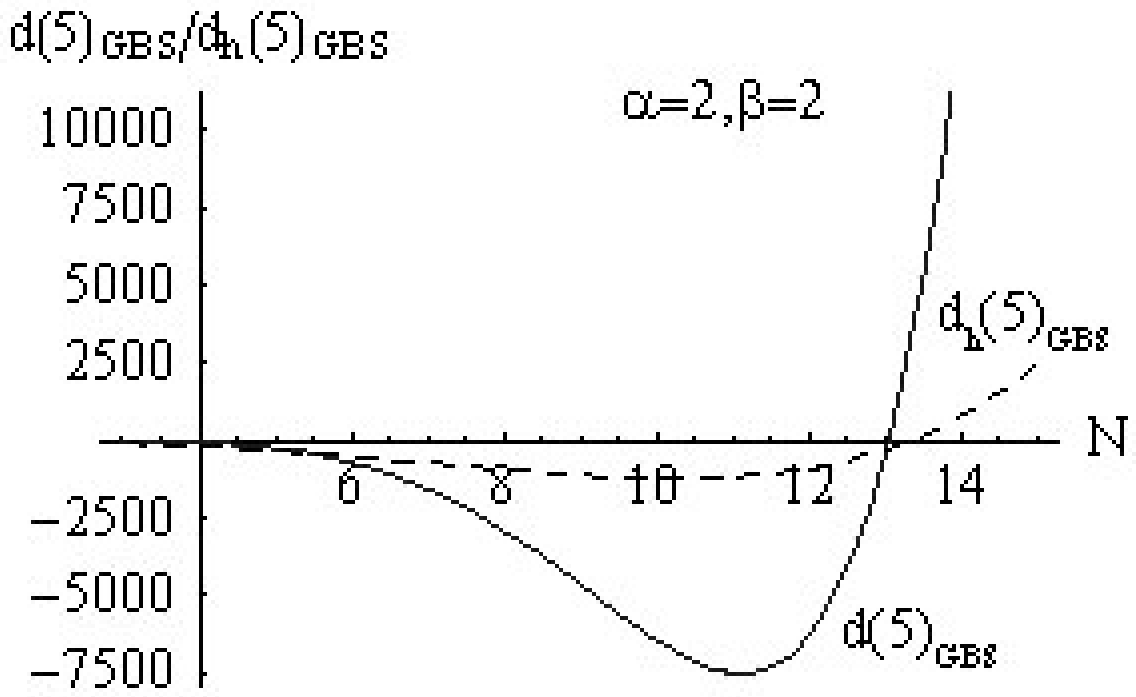

Figure 1: Shows 5th order HOA and HOSPS in GBS. Values of N, beyond which d(5) and $d_h(5)$ terminated are 13.05 and 13.35 respectively.

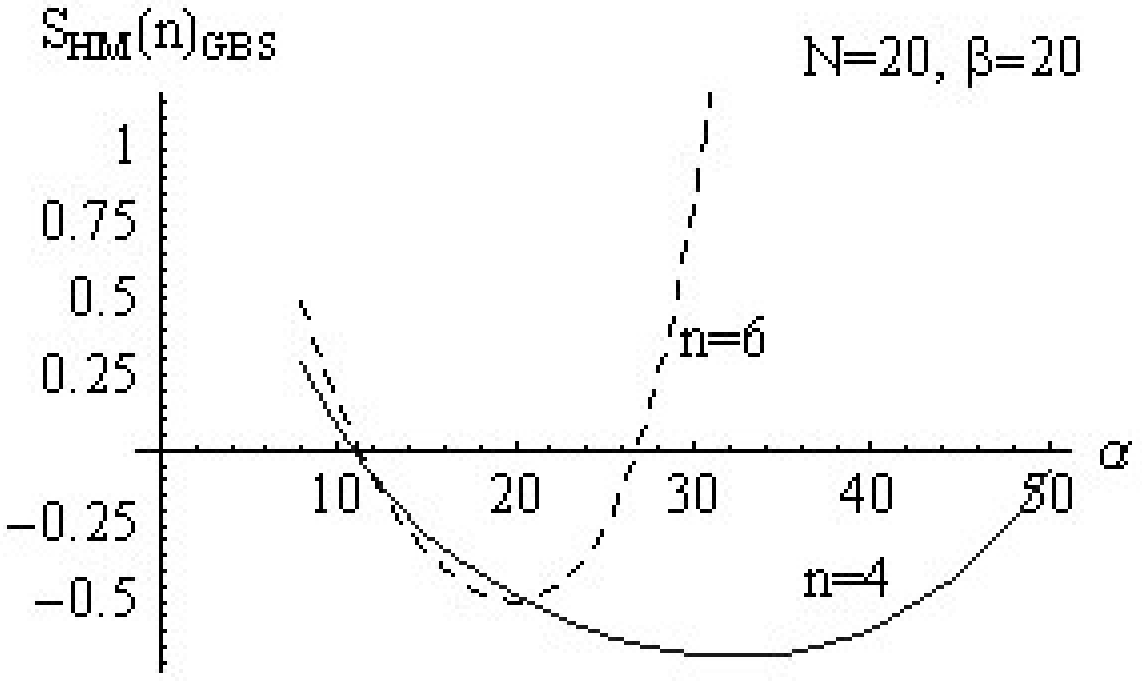

Figure 2: Shows 4th order and 6th order HOS with $\alpha$ in GBS.

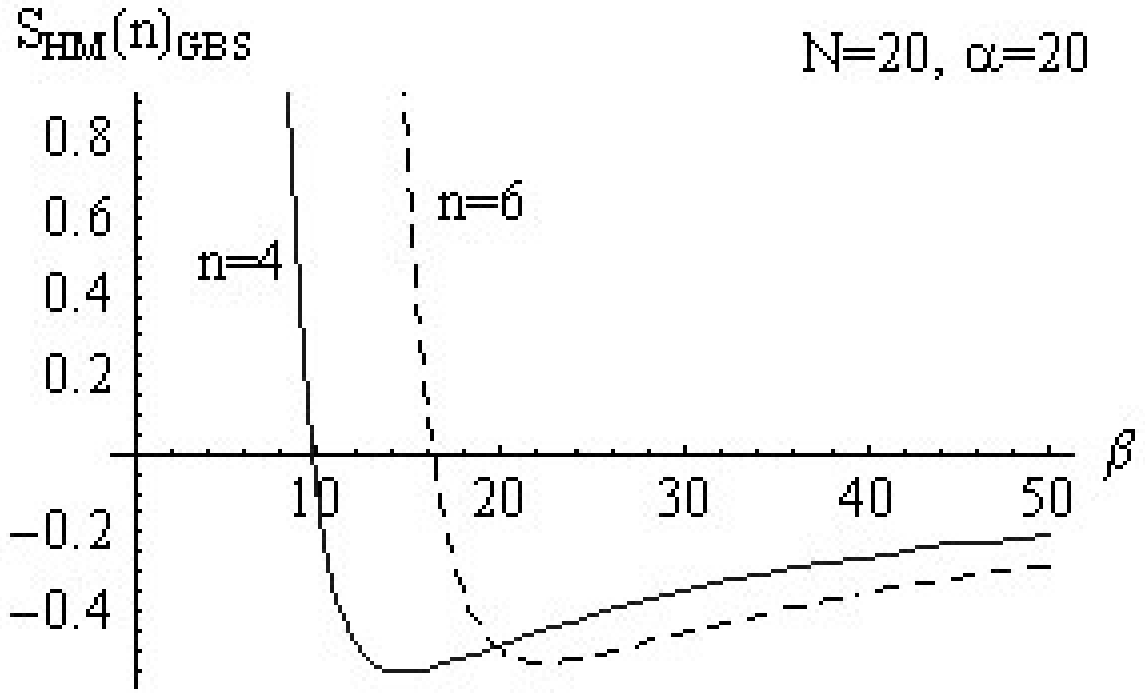

Figure 3: Shows 4th order and 6th order HOS with Beta in GBS. Possibility of HOS is high for large values of $\beta$ and can not have any upper limit.

| Parameters | d(3) | d(5) | dh(3) | dh(5) | SHM(4) | SHM(6) |
|---|---|---|---|---|---|---|
| $\alpha = 5$, $\beta = 30$ & $N = 30$ | 346.68 | 2.4x$10^4$ | 97.69 | 6.4x$10^3$ | 1.18 | 2.16 |
| $\alpha = 30$, $\beta = 5$ & $N = 30$ | -6.5x$10^4$ | -9x$10^7$ | -1.7x$10^3$ | -2.4x$10^5$ | 39.54 | 2724.10 |
| $\alpha = 5$, $\beta = 5$ & $N = 5$ | -27.98 | -244.14 | -14.96 | -361 | 1.39 | 20.03 |
| $\alpha = 30$, $\beta = 30$ & $N = 30$ | -5.57x$10^3$ | -3x$10^6$ | -3.46.43 | -3.9x$10^4$ | -0.47 | -0.59 |

Table 1: Numerical data for the existence of HOA, HOSPS and HOS in GBS.

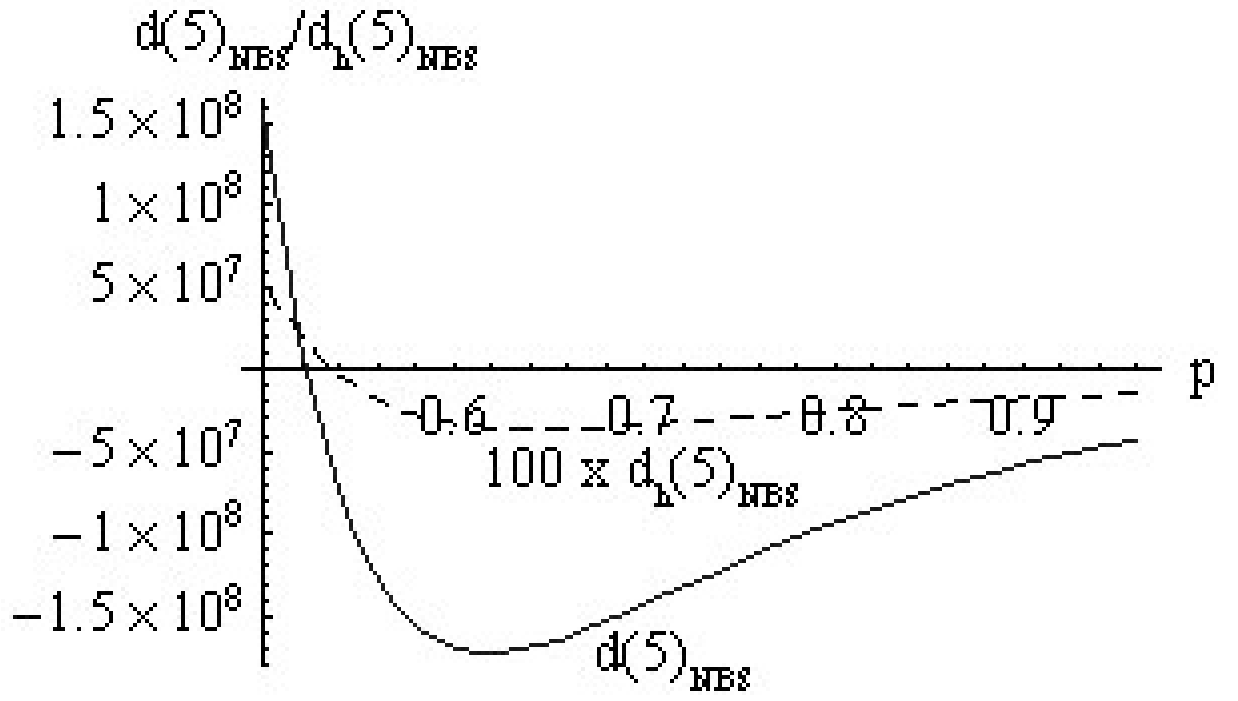

Figure 4: Shows 5th order HOA and HOSPS in NBS for M=20

From Fig. 1- Fig. 3 it is clear that the HOA, HOSPS and HOS can be observed in Roy and Roy generalized binomial state. It is easy to observe from Fig. 1- Fig. 3 that a particular kind of nonclassicality can be seen for particular values of $\alpha$, $\beta$ and $N$. It is also observed that the region of nonclassicality in HOS reduces with the increase in order of the nonclassicality [see Fig. 2]. Further it has been observed that for particular values of $\alpha, \beta$ and $N$ the state may show HOA and HOSPS in absence of HOS [see second and third row of Table 1]. from Table 1 it is also clear that the state does not show signature of any of these higher order nonclassicalities for $\alpha = 5$, $\beta = 30$ and $N = 30$. Thus we conclude that the possibility of observation of any particular kind of higher order nonclassicality and its depth depends on the choice of $\alpha$, $\beta$ and $N$.

## 4.2 Negative Binomial State (NBS)

Following Barnett [32] we can define Negative Binomial state (NBS) as

$$|\eta, M\rangle = \sum_{n'=M}^{\infty} C_{n'}(\eta, M)|n'\rangle \tag{23}$$

where $C_{n'}(\eta, M) = \left[\left(\begin{array}{c} n' \\ M \end{array}\right)\eta^{M+1}(1-\eta)^{n'-M}\right]^{1/2}$, $0 \leq \eta \leq 1$ and M is a non-negative integer. This intermediate state interpolates between number state and geometric state. Following the method adopted in the previous subsection, we can obtain

$$\langle a^{\dagger k} a^l \rangle = \left[\frac{p^{M+1}}{(1-p)^M}\right] \sum_{n'=M}^{\infty} \sum_{n'=M+l-k}^{\infty} \left[\binom{n'-l+k}{M}\binom{n'}{M}(1-p)^{2n'-l+k}\left\{\frac{n'!(n'-l+k)!}{(n'-l)!^2}\right\}\right]^{\frac{1}{2}} . \tag{24}$$

Hence,

$$\begin{array}{rcl} d_h(n-1)_{NBS} & = & \sum_{r=0}^{n}\sum_{k=1}^{r}\left[S_2(r,k)\,{}^nC_r(-1)^r\left\{\left[\frac{p^{M+1}}{(1-p)^M}\right]\sum_{n'=M}^{\infty}\binom{n'}{M}(1-p)^{n'}n'\right\}^{n-r},\right. \\ & & \left.\left\{\left[\frac{p^{M+1}}{(1-p)^M}\right]\sum_{n'=M}^{\infty}\binom{n'}{M}(1-p)^{n'}\frac{n'!}{(n'-k)!}-\left(\left[\frac{p^{M+1}}{(1-p)^M}\right]\sum_{n'=M}^{\infty}\binom{n'}{M}(1-p)^{n'}n'\right)^k\right\}\right] \end{array} \tag{25}$$

and

$$\begin{array}{rcl} S_{HM}(n)_{NBS} & = & \left[\frac{1}{\left(\frac{1}{2}\right)_{\frac{n}{2}}}\sum_{r=0}^{n}\sum_{i=0}^{\frac{r}{2}}\sum_{k=0}^{r-2i}(-1)^r\frac{1}{2^{\frac{n}{2}}}t_{2i}\,{}^{r-2i}C_k\,{}^nC_r\,{}^rC_{2i}\right. \\ & & \left\{2\left[\frac{p^{M+1}}{(1-p)^M}\right]\sum_{n'=M}^{\infty}\sum_{n'=M-1}^{\infty}\left[\binom{n'+1}{M}\binom{n'}{M}(1-p)^{2n'+1}(n'+1)\right]^{\frac{1}{2}}\right\}^{n-r} \\ & & \left[\frac{p^{M+1}}{(1-p)^M}\right]\sum_{n'=M}^{\infty}\sum_{n'=M+r-2i-2k}^{\infty}\left[\binom{n'-r+2i+2k}{M}\binom{n'}{M}(1-p)^{2n'-r+2i+2k}\right. \\ & & \left.\left.\frac{n'!(n'-r+2i+2k)!}{(n'-r+2i+k)!^2}\right]^{\frac{1}{2}}\right]-1\,. \end{array} \tag{26}$$

Fig. 4 and Fig. 5 clearly describes the variation of depth and region of HOA, HOSPS and HOS in NBS with different control parameters. It is clear from Fig 4 and Fig 5 that the higher order nonclassicality is not observed for lower values of $p$. It is also seen that the region of nonclassicality in HOS reduces with the increase in order of the nonclassicality [see Fig. 5].

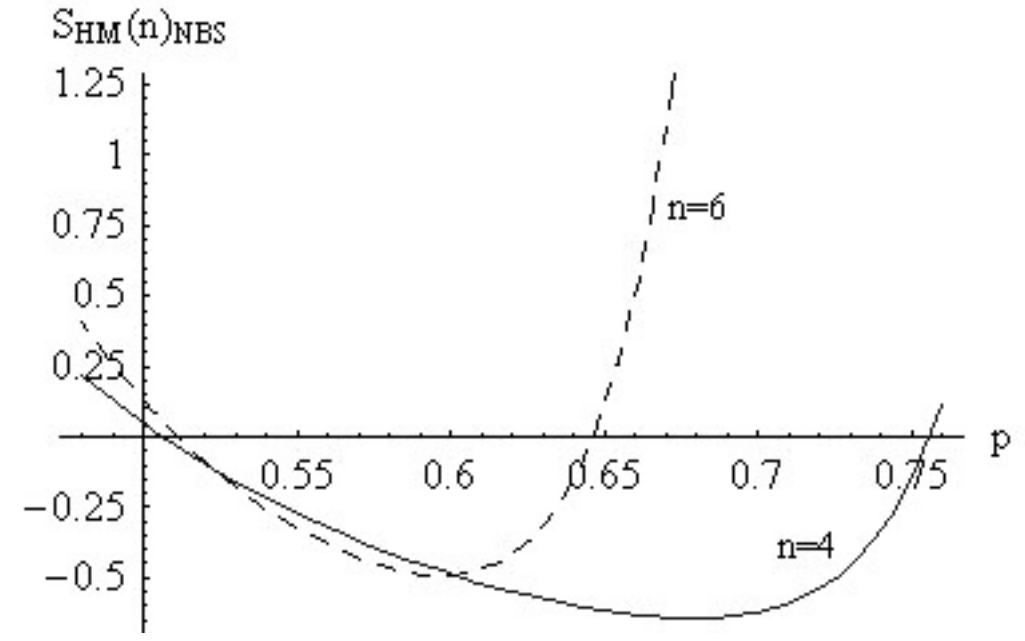

Figure 5: Higher order squeezing can be observed in NBS for M=20 and lies beyond p=0.5 only.

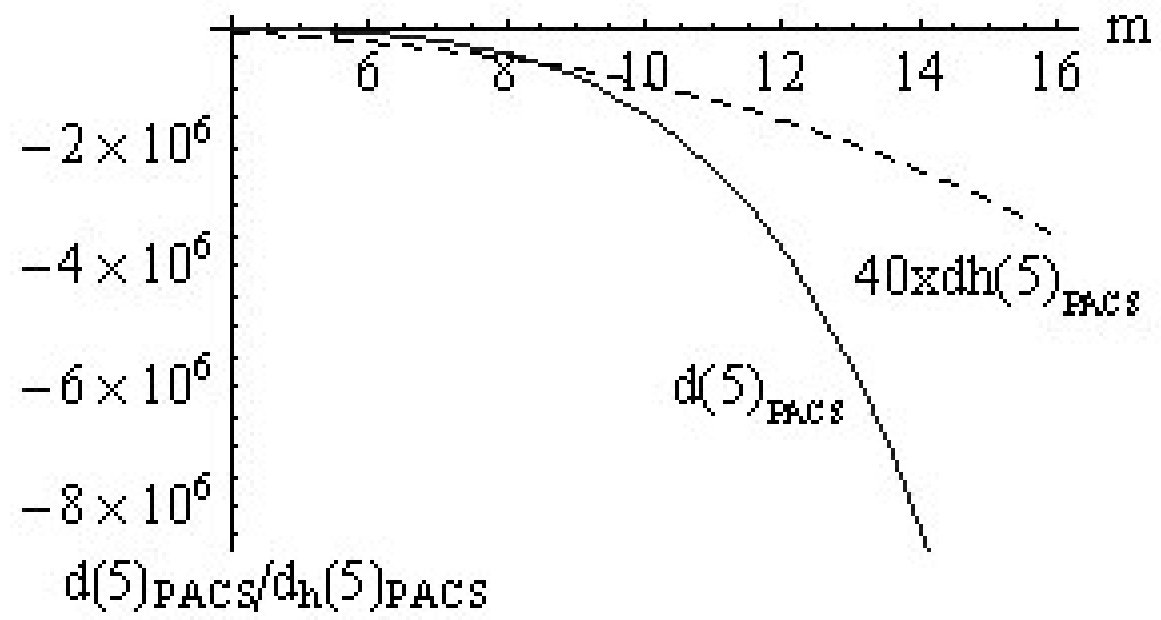

Figure 6: Shows 5th order HOA and HOSPS in PACS for $\alpha = 0.4$

### 4.3 Photon Added Coherent State (PACS)

Photon Added Coherent State is defined as [30]

$$|\alpha, m\rangle \quad = \quad N \sum_{n'=0}^{\infty} \frac{\alpha^{n}\sqrt{(m+n')!}}{n'!} |n' + m\rangle \tag{27}$$

where normalization constant $N = \frac{e^{-|\alpha|^2/2}}{\sqrt{L_m(-|\alpha|^2)m!}}$ and function $L_m(x) = \sum_{n'=0}^{m} \frac{(-x)^2 m!}{n'!^2 (m-n')!}$. Using the method adopted in the previous section we may obtain

$$\langle a^{\dagger k} a^l \rangle = N^2 \sum_{n'=0}^{\infty} \frac{\alpha^{2n'}}{n'!^2} \frac{(m + n' + k - l)!(n' + m)!}{(n' + m - l)!}. \tag{28}$$

Therefore, the parameters depicting signatures of HOSPS and HOS in PACS may be obtained as

$$\begin{array}{rcl} d_h(n-1)_{PACS} & = & \sum_{r=0}^{n} \sum_{k=1}^{r} \left[ S_2(r,k)\, {}^nC_r (-1)^r \left\{ N^2 \sum_{n'=0}^{\infty} \frac{\alpha^{2n'}}{n'!^2} \frac{(n'+m)!^2}{(n'+m-1)!} \right\}^{n-r} \right. \\ & & \left. \left\{ N^2 \sum_{n'=0}^{\infty} \frac{\alpha^{2n'}}{n'!^2} \frac{(n'+m)!^2}{(n'+m-k)!} - \left( N^2 \sum_{n'=0}^{\infty} \frac{\alpha^{2n'}}{n'!^2} \frac{(n'+m)!^2}{(n'+m-1)!} \right)^k \right\} \right] \end{array} \tag{29}$$

and

$$\begin{array}{rcl} S_{HM}(n)_{PACS} & = & \left[ \frac{1}{\left(\frac{1}{2}\right)_{\frac{n}{2}}} \sum_{r=0}^{n} \sum_{i=0}^{\frac{r}{2}} \sum_{k=0}^{r-2i} (-1)^r \frac{1}{2^{\frac{n}{2}}} t_{2i}\, {}^{r-2i}C_k\, {}^nC_r\, {}^rC_{2i} \right. \\ & & \left. \left\{ 2N^2 \sum_{n'=0}^{\infty} \frac{\alpha^{2n'}}{n'!^2} (n'+m)! \right\}^{n-r} \left\{ N^2 \sum_{n'=0}^{\infty} \frac{\alpha^{2n'}}{n'!^2} \frac{(m+n'-r+2i+2k)!(n'+m)!}{(n'+m-r+2i+k)!} \right\} \right] - 1 \end{array} \tag{30}$$

respectively. Variation of these parameters with $m$ are shown in Fig. 6 and Fig. 7. It is observed that HOA and HOSPS always exist in PACS but HOS is observed only for particular values of parameters as shown in Fig. 7. The results obtained in the present work is compared with the earlier results on PACS obtained by [6] and it is found that the results are in perfect coincidence.

### 4.4 Hypergeometric State (HS)

Hypergeometric state [27] is defined as

$$|L, M, \eta\rangle \quad = \quad \sum_{n'=0}^{M} H_{n'}^{M}(\eta, L)\, |n'\rangle \tag{31}$$

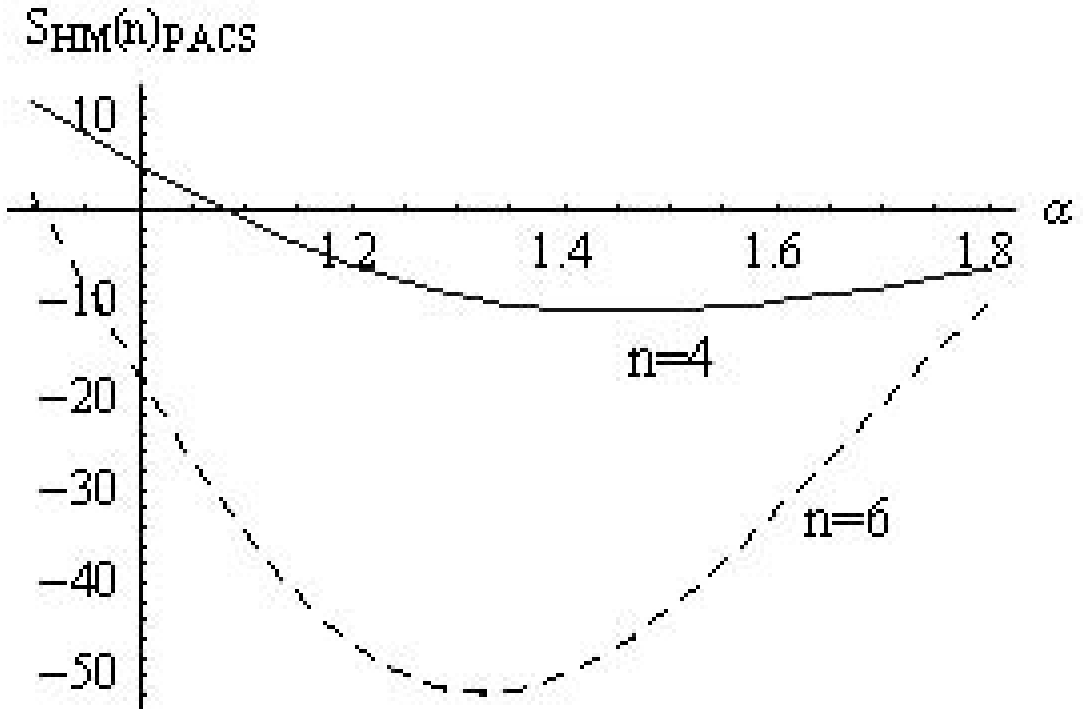

Figure 7: Higher order squeezing in PACS.

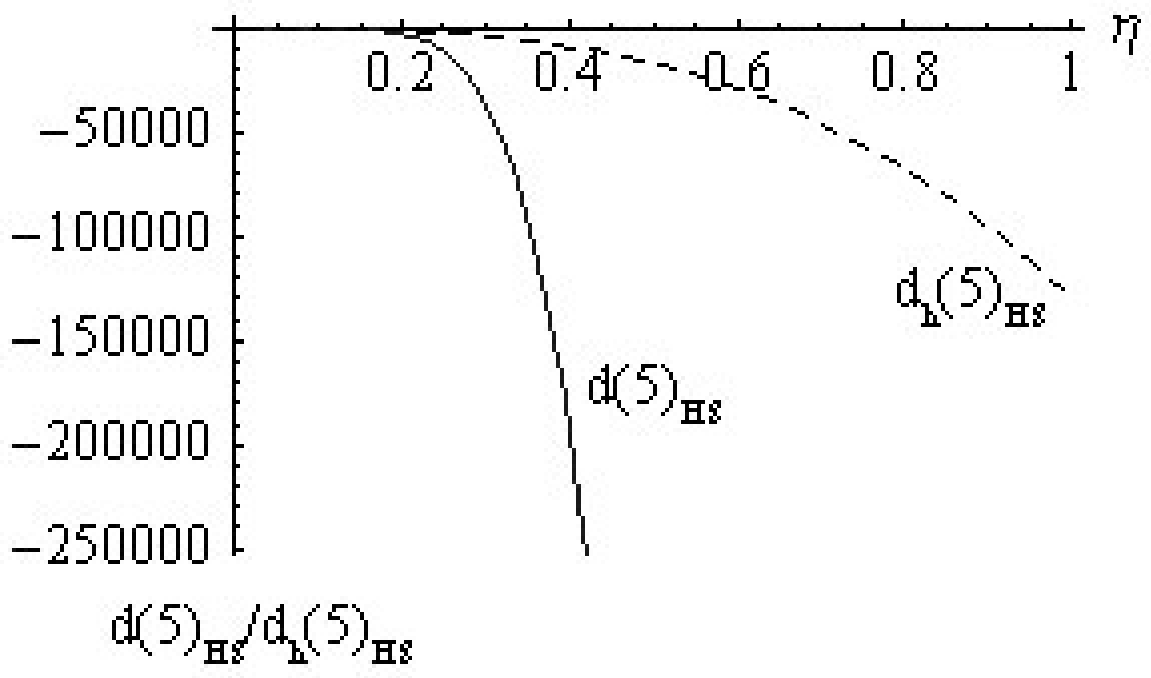

Figure 8: Shows 5th order HOA and HOSPS in HS

where $0<\eta<1$, L is a real number,

$$L \geqslant Max\left[\frac{M}{\eta},\frac{M}{1-\eta}\right] \tag{32}$$

and

$$H_{n'}^{M}(\eta,L) = \left[\binom{L\eta}{n'}\binom{L(1-\eta)}{M-n'}\right]^{1/2}\binom{L}{M}^{-1/2}. \tag{33}$$

Using this definition of HS and the procedure followed in 3.1, we obtain

$$\left\langle a^{\dagger k}a^{l}\right\rangle = \frac{M!(L\eta)!}{L!}\left[\frac{(L-k)!(L-l)!}{(M-k)!(M-l)!(L\eta-k)!(L\eta-l)!}\right]^{\frac{1}{2}}\sum_{n'=0}^{M-Max[k,l]}\left[\frac{\binom{L\eta-k}{n'}\binom{L(1-\eta)}{M-k-n'}\binom{L\eta-l}{n'}\binom{L(1-\eta)}{M-l-n'}}{\binom{L-k}{M-k}\binom{L-l}{M-l}}\right]^{\frac{1}{2}}. \tag{34}$$

Thus the analytic expression of the parameters that characterizes HOSPS and HOS may be obtained as

$$d_h(n-1)_{HS} = \sum_{r=0}^{n}\sum_{k=1}^{r}\left[S_2(r,k)\,{}^{n}C_r(-1)^r\left\{M\eta\right\}^{n-r}\left\{\frac{M!(L\eta)!(L-k)!}{L!(M-k)!(L\eta-k)!}-(M\eta)^k\right\}\right], \tag{35}$$

and

$$\begin{aligned} S_{HM}(n)_{HS} = {} & \frac{1}{\left(\frac{1}{2}\right)_{\frac{n}{2}}}\left[\sum_{r=0}^{n}\sum_{i=0}^{\frac{r}{2}}\sum_{k=0}^{r-2i}(-1)^r\frac{1}{2^{\frac{n}{2}}}t_{2i}\,{}^{r-2i}C_k\,{}^{n}C_r\,{}^{r}C_{2i}\right. \\ & \left[2\binom{L}{M}^{-1}(L\eta)^{1/2}\sum_{n'=0}^{M-1}\left[\binom{L\eta}{n'}\binom{L(1-\eta)}{M-n'}\binom{L\eta-1}{n'}\binom{L(1-\eta)}{M-1-n'}\right]^{1/2}\right]^{n-r} \\ & \frac{M!(L\eta)!}{L!}\left[\frac{(L-k)!(L-r+2i+k)!}{(M-k)!(M-r+2i+k)!(L\eta-k)!(L\eta-r+2i+k)!}\right]^{1/2} \\ & \left.\sum_{n'=0}^{M-Max[k,r-2i-k]}\left[\frac{\binom{L\eta-k}{n'}\binom{L(1-\eta)}{M-k-n'}\binom{L\eta-r+2i+k}{n'}\binom{L(1-\eta)}{M-r+2i+k-n'}}{\binom{L-k}{M-k}\binom{L-r+2i+k}{M-r+2i+k}}\right]^{\frac{1}{2}}\right]-1. \end{aligned} \tag{36}$$

Fig. 8 and Fig. 9 graphically represent (35) and (36). From these figures it is clear that the higher order nonclassicalities (HOSPS and HOS) can be observed in hypergeometric state. It is also observed that HOA and HOSPS always exist but HOS is not observed for higher values of $p$ . We have observed similar kind of behavior in BS too.

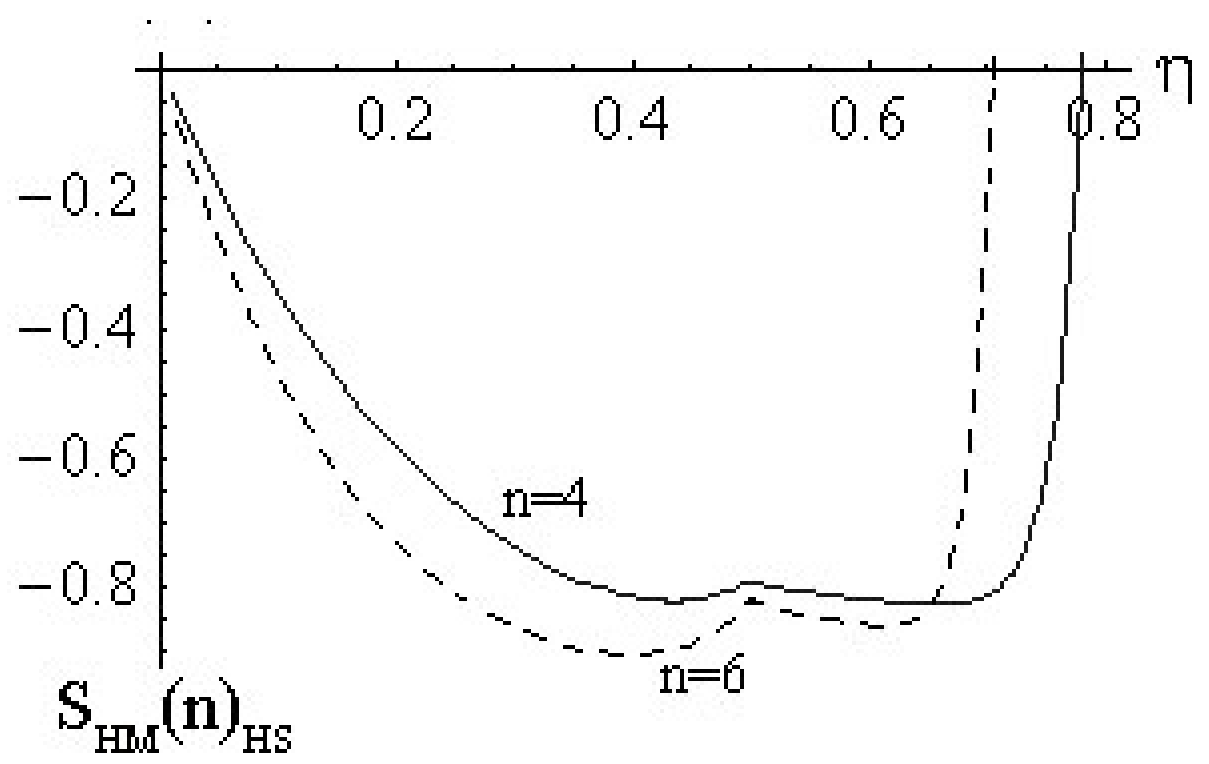

Figure 9: Shows 4th order and 6th order HOS in HS

# 5 Conclusions

In the present work, we have seen that the HOS and HOSPS can be observed in different intermediate states, e.g. GBS, NBS, PACS and HS. In GBS, it is observed that for particular values of $\alpha, \beta$ and $N$, it may show HOA and HOSPS in the absence of HOS [see Table 1]. It is also observed that the region of nonclassicality in HOS reduces with the increase in order of the nonclassicality. In NBS we have not observed higher order nonclassicality for lower values of $p$. It is also observed that HOA and HOSPS always exist in PACS but HOS is observed only for particular values of parameters. Similiarly, in hyper geometric state HOA and HOSPS always exist but HOS is not observed for higher values of $p$. In addition to these observations, earlier we have reported [1] that BS always shows HOA and HOSPS but it does not show HOS for all values of $p$. These observations clearly establish that HOA and HOS are mutually independent phenomenon. This is analogous to the corresponding observation in the lower order. Further, in [1] we had shown that NLVSS, which shows higher order squeezing does not show HOSPS and HOA. The opposite was observed in NLESS. With the help of some simple density matrices it was also shown in [1] that the HOA and HOSPS are independent of each other. Now we can combine all these observations and conclude that the existence of HOA, HOS and HOSPS are independent of each other. Therefore, to regrously study the higher order nonclassical properties of a quantum state one need to study HOA, HOS and HOSPS. But interestingly same experiment can detect the signature of all these nonclassical characteristics [1] of the radiation field. Finally we conclude that all the intermediate states studied here, confirms higher order nonclassical characteristics HOSPS and HOS in different regimes of parameters. Therefore appropriate choice of parameters is important to observe higher order nonclassicality in intermediate states. The simplified cirteria used here for the study of possibilities of observing HOS and HOSPS are also expected to be useful in the future to predict the existence of higher order nonclassicalities in other quantum states.

**Acknowledgment**: AP thanks to DST, India for partial financial support through the project grant SR\FTP\PS-13\2004.